\documentclass[aps,prl,preprint,showpacs,showkeys,groupedaddress,amsmath,amssymb]{revtex4}

\usepackage{graphicx}
\usepackage{dcolumn}
\usepackage{bm}

\begin{document}


\title{Nonresonant microwave absorption in epitaxial La-Sr-Mn-O films and its relation to colossal magnetoresistance}

\author{ M. Golosovsky\footnote{Permanent address: the Racah Institute of Physics, the Hebrew University of Jerusalem, 91904 Jerusalem, Israel},  P. Monod, }%
\affiliation{Laboratoire de Physique du Solide, ESPCI, 10 rue Vauquelin, 75231 Paris Cedex 05, France}
\author{P.K. Muduli and R.C. Budhani}%
\affiliation{Department of Physics,Indian Institute of Technology,Kanpur, 208016, India\\}
\author{L. Mechin and P. Perna\footnote{Permanent address: University of Cassino, Dipartimento di Meccanica Strutture Ambiente e Territorio, Facolta` di Ingegneria, via G. Di Biasio, 43, 03043 CASSINO (FR), Italy }}%
\affiliation{  GREYC (CNRS-UMR 6072),ENSICAEN and University of Caen,6 Blvd Marechal Juin-14050, Caen-Cedex, France
\\}
\date{\today} 
\begin{abstract} 
We study   magnetic-field-dependent nonresonant microwave absorption and dispersion in thin La$_{0.7}$Sr$_{0.3}$MnO$_{3}$ films  and show that it originates from the colossal magnetoresistance. We develop the model for  magnetoresistance of a thin ferromagnetic film in oblique magnetic field. The model accounts fairly well for our experimental findings, as well as for results of other researchers. We demonstrate that nonresonant microwave absorption is a powerful technique that allows contactless measurement of magnetic properties of thin films, including magnetoresistance, anisotropy field and coercive field.

\end{abstract}

\pacs {75.47.Lx, 75.47.Gk, 76.50.+g, 75.70.-i, 75.30.Gw}
\keywords{manganite, thin film, colossal magnetoresistance,  microwave absorption}
\maketitle
\section{introduction}
The colossal magnetoresistance of manganites has been extensively studied  by  transport methods and was  attributed to the double exchange  mechanism \cite{Salamon}, although the detailed mechanism  has not been  unambiguously established so far. This encourages  experimental study of  magnetoresistance in manganites by complementary methods. Contactless techniques, such as microwave absorption, are particularly advantageous here. This is the purpose of our present study- to explore potential of the microwave absorption technique to measure magnetoresistance in manganites. 

So far, microwave absorption in manganites was studied by the cavity perturbation technique. The measurements in constant field \cite{Tyagi,Robson,Owens}  revealed microwave absorption  linearly dependent  on magnetic field that was attributed to colossal magnetoresistance. The measurements in alternating field, using the field modulation technique \cite{Lofland,Budak,Lyfar}, revealed a very broad nonresonant absorption. Lyfar et al. \cite{Lyfar} assumed that it originates from the colossal magnetoresistance as well. In this work we systematically  study  the field and angular dependence of  nonresonant microwave absorption in epitaxial La$_{0.7}$Sr$_{0.3}$MnO$_{3}$ films and prove its magnetoresistive origin. 

We perform our measurements using a Bruker X-band  ESR spectrometer equipped with the bipolar current source. While such spectrometers are widely used  for magnetic resonance measurements,  their employment for  the study of nonresonant  microwave absorption was  restricted mostly to superconductors \cite{Enriquez}. In this work we develop a methodology to analyze the nonresonant microwave absorption measured using ESR spectrometer and field modulation technique.  Since ESR technique is very sensitive and versatile, our approach  opens a powerful opportunity to study  magnetoresistance of different materials in a contactless way.
\section{\label{sec:level1}Model}
\subsection{\label{sec:level2}A ferromagnetic film in oblique magnetic field- magnetostatics}
To calculate  magnetization of a thin ferromagnetic  film in oblique magnetic field we  consider its free energy:
\begin{equation}
\Phi=\Phi_{0} (M) + U_{anisotropy}+U_{demagnetization}+U_{Zeeman}
\label{energy}
\end{equation}
The first term here absorbs all  angular-independent contributions. We approximate it by $\Phi_{0}(M)=\frac{(M-M_{0})^{2}}{2\chi_{0}}+\emph{const}$, where  $M_{0}$ is the zero-field magnetization and $\chi_{0}=\frac{d M}{d H}$ is the bulk magnetic susceptibility.  We assume the "easy plane" anisotropy  (this includes the shape and the crystalline anisotropy), hence  $U_{anisotropy}+U_{demagnetization}=\frac{\beta M^{2}}{2}\cos^{2} \Theta$,  where  $\Theta$ is the polar angle of magnetization,  $\Psi$ is the polar angle of the external field, and $\beta>0$. The anisotropy field is $H_{a}=\beta M$.  Equation (\ref{energy}) reads then:
\begin{equation}
\Phi(M,\Theta)=\frac{(M-M_{0})^{2}}{2\chi_{0}} + \frac{\beta M^{2}}{2}\cos^{2} \Theta-MH\cos (\Theta-\Psi) \label{energy-magnetization}
\end{equation}
where we retain  only those terms that depend on $M$ and $\Theta$.  The equilibrium conditions $\left(\frac{\delta\Phi}{\delta \Theta}\right)_{M}=0$;  $\left(\frac{\delta\Phi}{\delta M}\right)_{\Theta}=0$  yield
\begin{subequations}
\label{eq:whole}
\begin{equation}
\beta M\sin \Theta \cos\Theta=H\sin({\Theta-\Psi})
\label{subeq:angle}
\end{equation}
\begin{equation}
\frac{M-M_{0}}{\chi_{0}}
+ \beta M\cos^{2} \Theta-H\cos (\Theta-\Psi)=0
\label{subeq:magnetization}
\end{equation}
\end{subequations}
To analyze the induced magnetization, $\Delta M=M(H)-M_{0}$, we reduce the above  equations  to a single one: 
\begin{equation}
\Delta M=\chi_{0}[H\cos(\Theta-\Psi)-H_{a}\cos^{2}\Theta]
\label{energy-magn}
\end{equation}
The two terms in square brackets account for the effects of the external and  anisotropy field, correspondingly. The analytical solution of Eq. (\ref{eq:whole}) is cumbersome, hence we consider simplifications which occur in extreme cases. 
\begin{itemize}
\item{}Parallel  orientation, $\Psi=\pi/2$. 
\begin{equation}
\Theta=\pi/2, \chi=\chi_{0}, \Delta M_{parallel}=\chi_{0}H
\label{parallel}
\end{equation}
\item{}Perpendicular  orientation, $\Psi=0$. 
\begin{itemize}
\item{}Low field, $H<H_{a}$.  
\begin{equation}
\cos\Theta=\frac{H}{H_{a}},\chi=0,\Delta M_{perp}^{low}=0
\label{perp-low} 
\end{equation}
Although magnetization is not orthogonal to the external field, the induced magnetization is zero. Indeed, the induced magnetization is determined by the internal rather than by the external field. In the perpendicular orientation, the internal field in exactly zero for $H<H_{a}$. 
\item{}High field , $H>H_{a}$. 
\begin{equation}
\Theta=0, \chi=\chi_{0},\Delta M_{perp}^{high}\approx \chi_{0}(H-H_{a})
\label{perp}
\end{equation}
\end{itemize}
\item{}Intermediate angles. 
\begin{itemize}
\item{}Low field, $H<<H_{a}$. 

Since magnetization is almost parallel to the film, then
\begin{equation}
\Theta\approx \pi/2,\Delta M=\chi_{0}H\sin\Psi.
\label{int-low}
\end{equation}
The susceptibility, $\chi(H,\Theta)=\frac{\partial M}{\partial H}$, exhibits discontinuity at zero field,
\begin{equation}
\chi_{+}-\chi_{-}=2\chi_{0}|\sin\Psi| \label{discontinuity}
\end{equation}
\item{}High field, $H>>H_{a}$. 

Since magnetization is almost collinear with the field, then
\begin{equation}
\Theta\approx\Psi+\frac{H_{a}\sin \Psi\cos \Psi}{H-H_{a}\cos2\Psi}\label{angle-approx}
\end{equation}
\begin{equation}
\Delta M\approx \chi_{0}H\left(1-\frac{H_{a}\cos^{2}\Psi}{H+H_{a}\sin^{2}\Psi}\right)\label{high-field}
\end{equation}
\begin{equation}
\frac{\chi}{\chi_{0}}=1-\left(\frac{H_{a}\sin \Psi \cos \Psi}{H+H_{a}\sin^{2}\Psi}\right)^{2}.\label{high-field1}
\end{equation}
\end{itemize}
\end{itemize}

\subsection{\label{sec:level3}A ferromagnetic film in oblique field-magnetoconductance}
We wish to calculate  conductivity of a thin ferromagnetic film in oblique magnetic field  having in mind manganite compounds.  Magnetic field is responsible for the colossal and anisotropic magnetoconductance.

\textit{Colossal magnetoconductance} in manganites is usually attributed to the double-exchange mechanism. According to this scenario, the conductivity  is determined by the magnitude of magnetization and is almost independent of its orientation  with respect  to crystallographic axes \cite{Salamon,Ziese,Favre,Wang}, in other words, $\sigma=\sigma (M^{2})$ \cite{Hundley,Tokura}. We consider the temperatures not too close to $T_{C}$, and  moderate fields, $H<$1 T,  when magnetoconductance  is  small compared to the zero-field conductance \cite{Li,Mathieu,Reversat}. Then, the relation between magnetoconductance, $\Delta\sigma(H)=\sigma(H)-\sigma_{0}$, and magnetoresistance,  $\Delta\rho(H)=\rho(H)-\rho_{0}$,  is especially simple: $\frac{\Delta\rho(H)}{\rho_{0}}= -\frac{\Delta\sigma(H)}{\sigma_{0}}$, where $\rho_{0}$ and $\sigma_{0}$ are zero-field resistivity/conductivity. From now on we focus on magnetoconductance and write
\begin{equation}
\sigma (H)\approx \sigma_{0}+\left(\frac{d\sigma}{d M}\right)_{H=0} \Delta M(H)
\label{CMR}
\end{equation}
We substitute Eq. (\ref{energy-magn}) into Eq. (\ref{CMR}) and find
\begin{equation}
\Delta\sigma^{CMR}=\sigma(H)-\sigma_{0}\approx\chi_{0} \frac{d\sigma}{d M}\left[H\cos(\Theta-\Psi)-H_{a}\cos^{2}\Theta\right]
\label{CMR1}
\end{equation}
Equation (\ref{CMR1}) depicts positive magnetoconductance whose magnitude is determined by the factor $\chi_{0}\frac{d\sigma}{d M}$ (here, $\chi_{0}$  is the ionic part of the magnetic susceptibility in the ferromagnetic state that arises from incomplete spin polarization of Mn-ions at finite temperature  and does not include Pauli susceptibility of charge carriers) and whose field and angular dependencies are determined by the terms in  square brackets. Equation \ref{CMR1} improves the previous result of O'Donnell et al. \cite{O'Donnell} who didn't consider the term $H_{a}\cos^{2}\Theta$ arising from the demagnetizing and anisotropy fields.

\textit{Anisotropic magnetoconductance}  can be written as follows \cite{O'Donnell}:
\begin{equation}
\sigma^{AMR}= (1-b\sin^{2}\Theta_{J})\sigma_{0}
\label{AMR}
\end{equation}
where  $\sigma_{0}$ is the (field-dependent) conductivity  when the current is parallel to magnetization, $\Theta_{J}$ is the angle between the magnetization and the current, and $b$  is a small dimensionless constant  which has positive sign in manganites \cite{Favre,Ziese}. In the common  magnetic resonance geometry  the microwave current is predominantly oriented  perpendicularly to the rotation axis (Fig. \ref{fig:setup}).  Then for negligible in-plane anisotropy and for oblique field orientation  we find $\Theta_{J}\approx 90^{0}-\Theta$. The conductivity in magnetic field is
\begin{equation}
\sigma (H)= (1-b\cos^{2}\Theta)\sigma_{0}
\label{AMR1}
\end{equation}


The field dependence of the conductivity arises from the term $\sigma_{0}(H)$- this is due to colossal magnetoresistance - and, since the angle $\Theta$ can be  field-dependent, - from the term $b\cos^{2}\Theta$  that accounts for the anisotropic magnetoresitance. The CMR is a strong effect which appears at all field orientations and results in conductivity linearly increasing with field that comes to saturation  at high field on the order of a few Tesla. On the other hand, the AMR is a weak effect which is prominent only at oblique field orientation and it comes to saturation  in a relatively small field, $H\sim H_{a}$ (less than 0.5 T). Hence, at high field the CMR makes the dominant contribution, while at low field and oblique field orientation the CMR and AMR contributions may be comparable.


\subsection{\label{sec:level4}Nonresonant microwave absorption and magnetoresistance}
The inductive methods for measuring magnetoresistance are  advantageous since they are contactless and allow for the sample rotation in magnetic field. This is most easily realized using a cavity perturbation technique where conducting sample is mounted off-center the resonant cavity \cite{Tyagi} or in the antinode of the microwave electric field \cite{Robson,Srinivasu,Sarangi}. We consider here a different setup where the sample is mounted in the antinode of the microwave magnetic field. Although this corresponds to the electric field node,  Zhai et al. \cite{Zhai} showed that the conductivity may be measured in this configuration as well. We prefer this setup since it allows for the measurement of the resonant magnetic susceptibility.

For the qualitative analysis of our measurements we follow Ref. \cite{Geshkenbein} and consider the effective magnetic susceptibility of an infinite conducting film in the uniform parallel microwave field,
\begin{equation}
\chi_{mw}=(1+\chi_{int})\frac{(\sinh u +\sin u)+i(\sinh u -\sin u)}{u(\cosh u +\cos u)}\label{plate1}-1
\end{equation}
Here, $\chi_{int}$ is the intrinsic microwave magnetic susceptibility of the film,  $d$ is the film thickness,  $\delta=(2/\mu\omega\sigma)^{1/2}$ is the skin-depth, $\omega$ is the microwave frequency, $\mu=\mu_0(1+\chi_{int})$ is the intrinsic magnetic permeability, and  $u=d/\delta$. For a thin film, $u<<1$, Eq. (\ref{plate1}) reduces to
\begin{equation}
\chi_{mw}\approx\chi_{int}-\frac{u^4}{30}+i\frac{u^2}{6}
\label{plate2}
\end{equation}
The first term in Eq. (\ref{plate2}) corresponds to intrinsic microwave susceptibility. It is non-negligible only at the ferromagnetic resonance. The resonance field is found from the well-known expression \cite{Gurevich}
\begin{equation}
\left(\frac{\omega}{\gamma}\right)^{2}=
[H_{res}\cos(\Theta-\Psi)-H_{a}\cos^{2}\Theta]
\times[H_{res}\cos(\Theta-\Psi)-H_{a}\cos2\Theta]
\label{FMR}
\end{equation}

The second and third terms in Eq.(\ref{plate2}) account for the eddy-current contribution to the real and imaginary parts of magnetic susceptibility, correspondingly (-see also Ref. \cite{Dyson}). To estimate this nonresonant contribution we neglect $\chi_{int}$ in Eq. (\ref{plate2}) and find
\begin{subequations}
\label{eq:susc}
\begin{equation}
\chi"_{mw}\approx \frac{\mu_{0}\omega\sigma d^{2}}{12}
\label{subeq:chi2}
\end{equation}
\begin{equation}
\chi'_{mw}\approx \frac{\mu_{0}^{2}\omega^{2}\sigma^{2} d^{4}}{120}<<\chi"_{mw}
\label{subeq:chi1}
\end{equation}
\end{subequations}

Equation (\ref{subeq:chi1}) indicates that the eddy-current contribution to the real part of the thin-film susceptibility is negligible, while Eq. (\ref{subeq:chi2}) shows that the lossy part of the effective magnetic susceptibility of a thin film  is proportional to the conductivity (this is in contrast to thick films where it is  proportional to the resistivity -Ref. \cite{Zhai}). Therefore, by measuring the $\chi"_{mw}$ of a thin film one can find conductivity.

\section{Experimental}
\subsection{The samples}
We studied  thin epitaxial La$_{0.7}$Sr$_{0.3}$MnO$_{3}$ films ($d=$ 50, 100, 150 and 200 nm)  on the (001)  SrTiO$_{3}$ substrate  and La$_{0.67}$Sr$_{0.33}$MnO$_{3}$ films ($d=$ 50 and 150 nm) on the NdGaO$_{3}$ substrate. The films were grown by the pulsed laser deposition technique in two different laboratories \cite{Budhani,Mechin} and cut to small $1\times1$ mm$^2$  pieces in order to keep the reasonable value of the cavity $Q$-factor. The film thickness is much smaller than the skin-depth at 10 GHz ($\delta=$22 $\mu$m at 295 K and $\delta=$5 $\mu$m at 50 K). Although the $T_{C}$ and coercive field  of the samples  were different, almost all of them showed measurable microwave magnetoconductance. Most part of the results shown here were obtained using 200 nm thick La$_{0.7}$Sr$_{0.3}$MnO$_{3}$ films on  SrTiO$_{3}$.

\subsection{The basics of the ESR spectrometer}
We utilized a bipolar Bruker ESR spectrometer operating at 9.4 GHz, a  $TE_{102}$ resonant cavity, and an Oxford cryostat. The sample is mounted in the cavity center. The microwave bridge measures reflected signal from the cavity with the sample. The bridge is balanced and  the cavity  is critically coupled at $H=$0. The  dc magnetic field is then slowly swept.   If the effective microwave magnetic susceptibility of the sample, $\chi_{mw}$, depends on magnetic field, the condition of critical coupling is  violated and a reflected signal appears:
\begin{equation}
P\propto \chi_{mw}(H) \eta Q\label{sample}
\end{equation}
Here $Q$ is the $Q$-factor of the cavity, $\eta\approx V/V_{c}$ is the filling factor, $V$ is the sample volume and $V_{c}$ is the cavity volume. The phase setting of the microwave detector chooses either absorption, $P_{abs}\propto \chi"_{mw}$, or dispersion, $P_{disp}\propto\chi'_{mw}$. To achieve high sensitivity, the dc field is modulated and the modulated reflection,
\begin {equation}
S_{mod}=\frac{dP}{dH}H_{mod}
\label{modulation}
\end{equation}
is  measured using a lock-in detector. Here, $H_{mod}$ is the amplitude of the modulation field. To find the absolute reflectivity, the modulated reflection is integrated,
\begin{equation} 
P(H)=P(H_{0})+\frac{1}{H_{mod}}\int^{H}_{H_{0}}{S_{mod}dH}
\label{integrated}
\end{equation}
Here $H_{0}$ is the lower  limit of the dc field sweep and $P(H_{0})$ should be determined independently.

\subsection {Extraction of the genuine signal} 
In the context of the ESR spectroscopy, the  nonresonant microwave absorption is represented by the modulation signal baseline that consists of several contributions: 
\begin{equation}
S_{mod}=S_{bridge}+S_{vibr}+S_{sample}
\label{S}
\end{equation}
\begin{itemize}
\item{$S_{bridge}$}, a constant offset that originates from the electronics of the microwave bridge. We use symmetric (with respect to the zero) field sweep, find the average signal $S_{ave}$, and subtract it from our results.  Since $S_{vibr}$ and $S_{sample}$ are odd functions of the field,  they do not contribute to the average signal, hence $S_{bridge}=S_{ave}$.

\item{$S_{vibr}\propto I_{mod}H$}. This contribution comes from the cavity wall vibration. Indeed, when the ac current $I_{mod}$ flows through the modulation coils (they are usually firmly attached to cavity walls) the cavity acquires periodic deformation in the dc magnetic field $H$. The shape of the cavity  and its  resonant frequency are  modulated, the condition of critical coupling is periodically violated and there appears a  reflected signal that is proportional to the mechanical force on cavity walls. This parasitic contribution linearly depends on field and can be easily taken for the magnetoresistance signal.

To eliminate vibration contribution, $S_{vibr}$, we choose a special phase setting of the lock-in detector.  Indeed, since the vibration signal  is  phase shifted with respect to the modulation current (most probably due to proximity of the modulation frequency to  mechanical resonances of the cavity) it is maximized at certain  phase setting of the lock-in detector, $\psi\neq 0$. However, the genuine signal, arising from magnetoconductance in the sample,  is in phase with the modulation field and it is maximized when   $\psi= 0$. Therefore, by setting the lock-in detector phase in quadrature with the vibration signal we eliminate the latter. Although the magnitude of the genuine signal  is also diminished by $\cos\psi$, it does not disappear completely. 

To verify the elimination of the vibration contribution we analyze the dispersion signal. Note, that vibration contributions into absorption and dispersion signals are comparable, while magnetoconductance contributes mostly to the absorption signal [Eq. (\ref{eq:susc})]. Therefore, the test for the proper choice of the lock-in detector phase  is  that the linear baselines in derivative dispersion and absorption  signals disappear simultaneously.
 
\item{$S_{sample}$} is the genuine signal which arises from the microwave absorption in the sample. The nonresonant absorption signal associated with the conductor loss is found from Eqs. (\ref{subeq:chi2},\ref{modulation}).
\begin{equation}
S_{sample}=\frac{d\sigma}{d H}\frac{\mu_{0}\omega d^{2}\eta QH_{mod}}{12}
\label{measurement}
\end{equation}
It consists of the field derivative of the conductivity, $\frac{d\sigma}{d H}$, multiplied by a constant factor. To find  conductivity we integrate Eq. (\ref{measurement}). The results are analyzed  using Eqs. (\ref{eq:whole},\ref{CMR1}).
\end{itemize}

\section{Experimental results and comparison to the model}
\subsection{Derivative microwave absorption}
Figure \ref {fig:derivative} shows derivative microwave absorption at several field orientations for a La$_{0.7}$Sr$_{0.3}$MnO$_{3}$ film on the SrTiO$_{3}$ substrate \cite{Budhani}.  We observe a sharp FMR signal superimposed on a  wide antisymmetric  baseline. For the parallel field orientation  the baseline may be represented as a step function which  reverses its sign at zero field.  For oblique field orientation, the baseline varies with field more gradually and the zero-field discontinuity becomes smaller. When the field deviation  from the perpendicular orientation  is less  than  10$^0$, there appears  pronounced zero-field absorption associated with  magnetic domains. It will be discussed elsewhere \cite{we-domain}. In this study we focus on the baseline which we attribute to microwave magnetoconductance.

The model prediction  based on Eqs. (\ref{subeq:angle},\ref{subeq:magnetization},\ref{CMR1}) fits the baseline  fairly well (see solid lines in the Fig. \ref{fig:derivative}). For the whole family of $\frac{dP(H,\Theta)}{dH}$  dependencies there are only two fitting parameters:  the high-field absorption derivative in the parallel orientation  and the effective anisotropy field, $H_{a}^{MR}=$3900 Oe. The latter is somewhat lower than the anisotropy field $H_{a}^{FMR}=$ 4300 Oe found from the ferromagnetic resonance  using Eq.(\ref{FMR}) and Fig. \ref{fig:derivative}.  The difference may arise from the variation of the anisotropy field across the film.

Figure \ref{fig:model} shows angular dependence of the derivative microwave absorption at high field, $H=$1 T. This very weak dependence  is consistent with the model prediction [Eq. (\ref{high-field1})].

\subsection{Integrated microwave absorption}
To compare integrated absorption at different orientations we integrated the data of the  Fig. \ref{fig:derivative} using  Eq. (\ref{integrated}). The main difficulty here is to find $P(H_{0})$ [Eq. (\ref{integrated})] which can depend on orientation. We performed integration  with arbitrary reference field and plotted the resulting curves for all orientations at the same plot. Different choices of $P(H_{0})$  correspond to  vertical shifts of the  curves [Eq. (\ref{integrated})]. We shift each curve vertically to achieve the same value of the minimal absorption. 
Figure \ref{fig:integrated} shows our results. We  observe  (i) superlinear field dependencies which we attribute to magnetoconductance and (ii) a small high-field bump arising from the ferromagnetic resonance. The superposition of both kinds of absorption in the same experimental run unambiguously proves the positive sign of magnetoconductance in LSMO.  Figure \ref{fig:model} shows angular dependence of the integrated absorption  at high field. The experimental data agree well with the model  prediction [Eqs. (\ref{high-field})] and this is an additional proof that the nonresonant microwave absorption arises from magnetoconductance. 

We consider now the curve for the almost perpendicular orientation, $\Psi=8^0$. When $H<$4000 Oe, the CMR contribution is negligible since the internal field is zero [Eq. (\ref {perp-low})]. However, there is some low-field absorption which corresponds to negative magnetoconductance.  Similar feature was observed in dc-transport measurements in LCMO films \cite{Eckstein} and was attributed to anisotropic magnetoresistance. The AMR can explain our results as well, although ferromagnetic resonance in the multidomain state \cite{Vucadinovic} seems to be a more plausible explanation here.

\subsection{Dispersion}
Figure \ref{fig:abs-disp-int} compares integrated absorption and dispersion for nearly parallel orientation.  With respect to the ferromagnetic resonance contribution, it is clearly seen that the magnitude of the  absorption peak at $H_{FMR}=2000$ Oe is equal to the peak-to-peak magnitude of the  dispersion peak, as expected for the Lorentzian resonance. 

With respect to the nonresonant eddy-current contribution, it appears only in the absorption signal and is absent in the dispersion signal. This conclusion is further verified by the inspection of the low-field derivative signals  which are measured with higher sensitivity.  Figure \ref{fig:abs-disp-der} shows that  absorption derivative exhibits a sharp zero-field discontinuity (this is  a signature of magnetoresistance- see Eq. (\ref{discontinuity})) while the dispersion derivative varies smoothly across zero, showing no trace of the magnetoresistance contribution. This is consistent with Eq.  (\ref{eq:susc}) that for our films yield $\chi`/\chi``=d^{2}/5\delta{^2}\sim 10^{-4}$.

\subsection{Temperature dependence}
To characterize microwave magnetoconductance at different temperatures we chose the factor $\frac{\partial P}{\partial H}$.  Equations (\ref{CMR1},\ref{measurement}) yield that in the parallel geometry it is field-independent and $\frac{\partial P}{\partial H} \propto
 \frac{\partial \sigma}{\partial M}$, hence, it is a direct measure of magnetoconductance.  Figure \ref{fig:MR-H-T} shows temperature dependence of $\frac{dP}{dH}$ as well as that of the anisotropy field. The latter was found from the ferromagnetic resonance data in the same sample. It should be noted that for our thin films the anisotropy field is dominated by the shape anisotropy and therefore, it can serve as a direct measure of magnetization.

Figure \ref{fig:MR-H-T} shows that  magnetoconductance  exhibits  a sharp peak around 300 K and becomes very small at low temperatures. This temperature dependence  is very similar to that for CMR found in  transport measurements on high-quality epitaxial La$_{0.7}$Sr$_{0.3}$MnO$_{3}$ films \cite{Li,Mathieu,Reversat} and indicates an intrinsic  mechanism.   The inset to Fig. \ref{fig:MR-H-T} shows how magnetoconductance depends on the anisotropy field (in other words, on magnetization). 

It should be noted that the low-temperature measurements  were performed with one sample while  the measurements above room temperature  were performed with another sample which was cut from the same film. While the low-temperature measurements were performed at several fixed temperatures and low microwave power to exclude self-heating, the measurements above 295 K were performed at  ambient temperature and increased microwave power. In this case the sample temperature is enhanced  due to self-heating. In these latter measurements we directly measure the magnetoconductance and the anisotropy field while the sample temperature  was estimated using the data from  the inset to the Fig. \ref{fig:MR-H-T} assuming linear temperature dependence of the anisotropy field in the vicinity of $T_{C}$. 

\subsection{Low-field range}
Figure \ref{fig:hysteresis}  shows low-field integrated absorption and absorption derivative  in the parallel orientation. There is a pronounced hysteresis. The absorption derivative exhibits discontinuity that corresponds to the cusp in the integrated absorption. When the field orientation deviates from the parallel orientation,  the magnitude of the discontinuity  becomes smaller (Fig. \ref{fig:derivative}) and its position moves to higher field (Fig. \ref{fig:peak-coerc}). To account for the angular dependence of the discontinuity we note that in low field, $H<<H_{a}$, where  magnetization is nearly parallel to the film, the discontinuity should occur when the parallel field component is equal to coercive field,   $H_{||}=H\sin\Psi=H_{c}$. Indeed, the angular dependence $H_{disc}=\frac{H_{c}}{\sin\Psi}$ describes our data perfectly well (the solid line in Fig. \ref{fig:peak-coerc}). 
We find $H_{c}=$11 Oe and this is in good agreement with the SQUID measurements on the same film. The $H_{c}$ found in our measurements  increases at low temperatures, ($H_{c}$=23 Oe at 219 K and $H_{c}$=50 Oe at 4.2 K) that is consistent with magnetic measurements  for similar films \cite{Lecoeur}.

The deviation of the low-field integrated absorption from the linear dependence predicted by Eq. (\ref{parallel}) (see low-field region in Fig. \ref{fig:hysteresis})  corresponds to conductance drop. Similar low-field features in the magnetically unsaturated state were observed  in microwave studies of thin Fe-Cr films \cite{Frait,Krebs,Dubowick} and in dc-transport measurements in La$_{0.7}$Ca$_{0.3}$MnO$_{3}$  \cite{O'Donnell1,O'Donnell} and La$_{0.85}$Sr$_{0.15}$MnO$_{3}$ thin films \cite{Klein}. The low-field conductance drop in the magnetically unsaturated state can be related to the (i) domain wall resistance \cite{Golosov}, (ii)  ferromagnetic resonance in the multidomain state \cite{Vucadinovic}, or (iii) anisotropic magnetoresistance \cite{O'Donnell1,O'Donnell,Klein}. We attribute the low-field absorption in the nearly perpendicular orientation, $\Psi<10^{0}$, to the ferromagnetic resonance in the multidomain state \cite{we-domain}, while the conductance drop observed at $\Psi>10^{0}$ (Fig. \ref{fig:hysteresis}) is attributed to anisotropic magnetoresistance. 

Indeed, at high field the film is in the  single domain state and the projection of magnetization on the film plane is parallel to the microwave current (Fig. \ref{fig:setup}), hence $\Theta_{J}=0$ and  Eq. (\ref{AMR}) yields $\sigma=\sigma_{0}$.  However, when the in-plane field is below the in-plane saturation field, $H<H_{sat}$, the magnetization aligns along the easy in-plane axes and is not parallel to the current anymore.  Equation (\ref{AMR1}) yields $\sigma= (1-b\overline{\sin^{2}\Theta_{J}})\sigma_{0}$.
Therefore, when the field becomes smaller than the in-plane saturation field, there is a conductance drop,  $\Delta\sigma\sim-b\sigma_{0}$, that does not depend  on the polar angle of the field, $\Psi$. This is in agreement with our observations (Figs. \ref{fig:hysteresis},\ref{fig:peak-coerc}). Following this interpetation, the onset for the deviation from  the linear dependence predicted  by Eq. (\ref{CMR1}) indicates the saturation field. Figure \ref{fig:hysteresis} yields $H_{sat}=$ 90-100 Oe in good agreement with  $H_{sat}=$90 Oe found in magnetization measurements on similar films \cite{Suzuki}. 

\subsection{Comparison between the samples}
Almost all our films demonstrated measurable magnetoconductance. The field and angular dependencies for the films fabricated in different laboratories were much more the same.  Indeed, Figs. \ref{fig:abs-disp-int},\ref{fig:G197-int}  show  microwave magnetoconductance for two epitaxial La$_{0.7}$Sr$_{0.3}$MnO$_{3}$ films  on the SrTiO$_{3}$  that were fabricated in different laboratories. The field dependence of the nonresonant microwave absorption is very similar, although resonant absorption is different. 

\section{Discussion}
We verified our model and demonstrated that magnetoresistance is determined by the dc magnetic susceptibility in the ferromagnetic state.  This offers  possibility of measuring magnetostatic properties of magnetic films using contactless microwave methods, as was suggested earlier in Ref. \cite{Prinz}. In such a way we were able to measure coercive field (Fig. \ref{fig:peak-coerc}), the in-plane saturation field, and the in-plane anisotropy field (not shown here). In what follows we estimate the out-of-plane anisotropy field from the nonresonant microwave absorption. Indeed, for the integrated microwave absorption in the perpendicular orientation, Eq. (\ref{perp}) yields superlinear field dependence with the horizontal intercept equal to the anisotropy field. We  extrapolate  our data (Fig. \ref{fig:integrated}) to low field and find $H_{a}^{MR}=$3900 Oe. This is reasonably compared to the anisotropy field found from the ferromagnetic resonance as given by Eq. (\ref{FMR}), namely, $H_{a}^{FMR}=$4300 Oe.

In what follows we apply our approach to the data of Liu and Furdyna \cite{Furdyna} who measured ferromagnetic resonance in magnetic semiconductors.  The inset in Fig. \ref{fig:Furdyna}  shows absorption derivative in the perpendicular geometry for a thin  GaMnAs film \cite{Furdyna}. There is a ferromagnetic resonance accompanied with several spin-wave resonances, and a broad baseline that Ref. \cite{Furdyna} attributed to magnetoresistance. We integrate these data assuming that absorption derivative for $H<$ 2 kOe is negligibly small. Figure \ref{fig:Furdyna} shows the field dependence of the integrated absorption found in such a way. It is very  similar to what is shown in  Fig. \ref{fig:integrated} for a manganite film. We approximate the high-field data of the Fig. \ref{fig:Furdyna} by a linear dependence. The horizontal intercept yields $H_{a}^{MR}=$3700 Oe while the FMR yields somewhat higher value, $H_{a}^{FMR}=$4300 Oe. The difference  may arise from slight deviation from the perpendicular orientation, mosaicity, the spread of the magnitude of the  anisotropy field across the film, and some ambiguity in the determination of the magnitude of the zero-field absorption. 

\section{Conclusions}
We demonstrate how to measure magnetoresistance  in a contactless way using a bipolar  ESR spectrometer. We develop a model accounting for the magnetoresistance of a thin ferromagnetic film in  oblique magnetic field. Our model is in excellent agreement with our measurements. We show that the intrinsic magnetoresistance in La$_{0.7}$Sr$_{0.3}$MnO$_{3}$ films is determined by the magnitude of magnetization and is  insensitive to its orientation with respect to the film and to the crystallographic axes. Our approach to measure magnetoresistance using a microwave technique can be useful for other conducting magnetic materials.  
\begin{acknowledgments}
We are grateful to Denis Golosov and Lior Klein for helpful discussions, and to Xiangzhen  Xu   for the help with the handling of the samples. 
\end{acknowledgments}

\pagebreak 

\pagebreak
\begin{figure}[ht]
\caption{Experimental setup. The sample, which is a thin film on substrate, is firmly attached to the sample holder and is mounted  in the center of the resonant cavity, in the antinode of microwave magnetic field. The dc magnetic field, $H$, is perpendicular to the microwave magnetic field, $h_{mw}$, and is oriented at an angle $\Psi$ with respect to the film normal. This angle can be varied by rotating the sample.} 
\label{fig:setup}
\end{figure}

\begin{figure}[ht]
\caption{Absorption derivative at different orientations of the magnetic field. 
$\Psi=0^{0}$ and $\Psi=90^{0}$  stand for the perpendicular and parallel orientation, correspondingly. The arrow shows direction of the field sweep. Modulation field is 10 Oe. The  experimental data is shown by small circles. The sharp peaks correspond to the ferromagnetic resonance, while the broad antisymmetric  baseline originates from the microwave magnetoconductance. The solid lines show model prediction for the baseline. Two  fitting parameters have been used for all curves: anisotropy field, $H_{a}=$3900 Oe,  and high-field absorption derivative in the parallel geometry, 
$\frac{dP}{dH}= 2.1\times 10^{4}$.}
\label{fig:derivative}
\end{figure}

\begin{figure}[ht]
\caption{Angular dependence of the integrated absorption (filled symbols) and absorption derivative (open symbols) at $H=$1 T   as inferred from   Figs. \ref{fig:derivative},\ref{fig:integrated}. The solid lines show model prediction based on Eqs. (\ref{high-field},\ref{high-field1}) and $H_{a}=$3900 Oe.}
\label{fig:model}
\end{figure}

\begin{figure}[ht]
\caption{Integrated absorption at different orientations  of the magnetic field as found  from the integration of the data of the Fig. \ref{fig:derivative}. The curves for different orientations are vertically shifted to achieve the same value of minimal absorption. The numbers at each curve show the polar angle of the field, whereas $\Psi=0^0$ and $\Psi=90^0$ stand for the perpendicular and  parallel orientation, correspondingly. The dashed line shows linear approximation for  the high-field data for (almost) perpendicular orientation. The horizontal intercept yields the anisotropy field $H_{a}$. }
\label{fig:integrated}
\end{figure}

\begin{figure}[ht]
\caption{Integrated absorption and dispersion in the parallel geometry for a 200 nm thick La$_{0.7}$Sr$_{0.3}$MnO$_{3}$ film on SrTiO$_{3}$  \cite{Budhani}. Absorption signal exhibits ferromagnetic resonance at 2000 Oe  superimposed on the broad baseline that linearly depends on field. Dispersion signal exhibits only ferromagnetic resonance and does not show any baseline.}
\label{fig:abs-disp-int}
\end{figure}

\begin{figure}[ht]
\caption{Absorption derivative and dispersion derivative  in the parallel geometry. Note zero-field discontinuity in absorption  as opposed to smooth variation of dispersion. The black arrow shows direction of the field sweep. Modulation field, $H_{mod}=$10 Oe, is  high to achieve enough sensitivity for dispersion measurement. This leads to some distortion in the absorption curve due to overmodulation.} 
\label{fig:abs-disp-der}
\end{figure}

\begin{figure}[ht]
\caption{Temperature dependence of  the effective anisotropy field, $H_{a}$, and of the magnetoconductance, $\frac{dP}{dH}$, at $H=$1 T. The inset shows dependence of the derivative absorption on the anisotropy field. Filled symbols stand for the sample LSMO55-1-A. Here, the measurements were performed at  several fixed temperatures and at low microwave power of -30 dB. Open symbols stand for the sample LSMO55-1-B (both samples were cut from the same film). Here, the measurements were performed at ambient temperature and at different microwave power levels from -30 dB to 0 dB. In this case the temperature of the sample increases due to self-heating. It was estimated indirectly from the position of the FMR peak.}
\label{fig:MR-H-T}
\end{figure}

\begin{figure}[ht]
\caption{Low-field absorption:  integrated (main panel) and derivative (inset). The direction of field sweep is shown by arrows. The derivative absorption exhibits low-field discontinuity that is preceded by a sharp dip/peak. The integrated absorption exhibits  the cusp and the drop $\Delta P$, correspondingly. The dashed line shows linear extrapolation of the high field data to the low field region. Deviation of the data from the linear dependence indicates the in-plane saturation field, $H_{sat}$. The width of the hysteresis loop yields coercive field, $H_{c}=$ 11 Oe. We used here a very slow sweep rate and small modulation field, $H_{mod}=$0.5 Oe, to prevent  overmodulation.}
\label{fig:hysteresis}
\end{figure}

\begin{figure}[ht]
\caption{Angular dependence of the discontinuity field and of the low-field  absorption drop, $\Delta P$ (from the data of Fig. \ref{fig:hysteresis}). The solid line shows model prediction while the dashed line is the guide to the eye. The $H_{disc}$ is strongly angular-dependent while The $\Delta P$ is not. We attribute $H_{disc}$ to the coercive field and $\Delta P$ to the anisotropic magnetoresistance in the magnetically unsaturated state.} 
\label{fig:peak-coerc}
\end{figure}
\begin{figure}[ht]
\caption{Integrated absorption and dispersion in the parallel geometry for a 
100 nm thick La$_{0.7}$Sr$_{0.3}$MnO$_{3}$ film on SrTiO$_{3}$ \cite{Mechin}. Note similarity to Fig. \ref{fig:abs-disp-int}. The film is slightly nonuniform and the FMR peak is split in two.}
\label{fig:G197-int}
\end{figure}

\begin{figure}[ht]
\caption{The inset shows derivative microwave absorption in the perpendicular geometry for a 200 nm thick Ga$_{0.924}$Mn$_{0.076}$As film (replotted using the data of Ref. \cite{Furdyna}). The main panel shows integrated absorption calculated using the data of the inset and assuming that the absorption derivative at $H<$ 2 kOe is negligibly small. The dashed line shows linear approximation for high field.  The intercept of this line with the horizontal axis yields anisotropy field, $H_{a}$=3700 Oe.}
\label{fig:Furdyna}
\end{figure}

\begin{thebibliography}{}
\bibitem{Dyson} The resonant susceptibility is also affected by the finite conductivity. This leads to Dysonian lineshape arising from the mixing of the dispersive and absorptive parts of the resonant susceptibility.
\end{thebibliography}
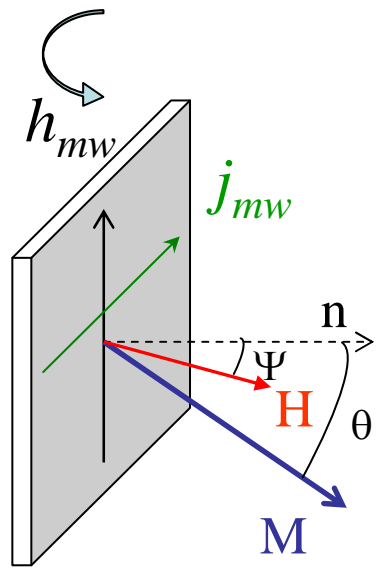
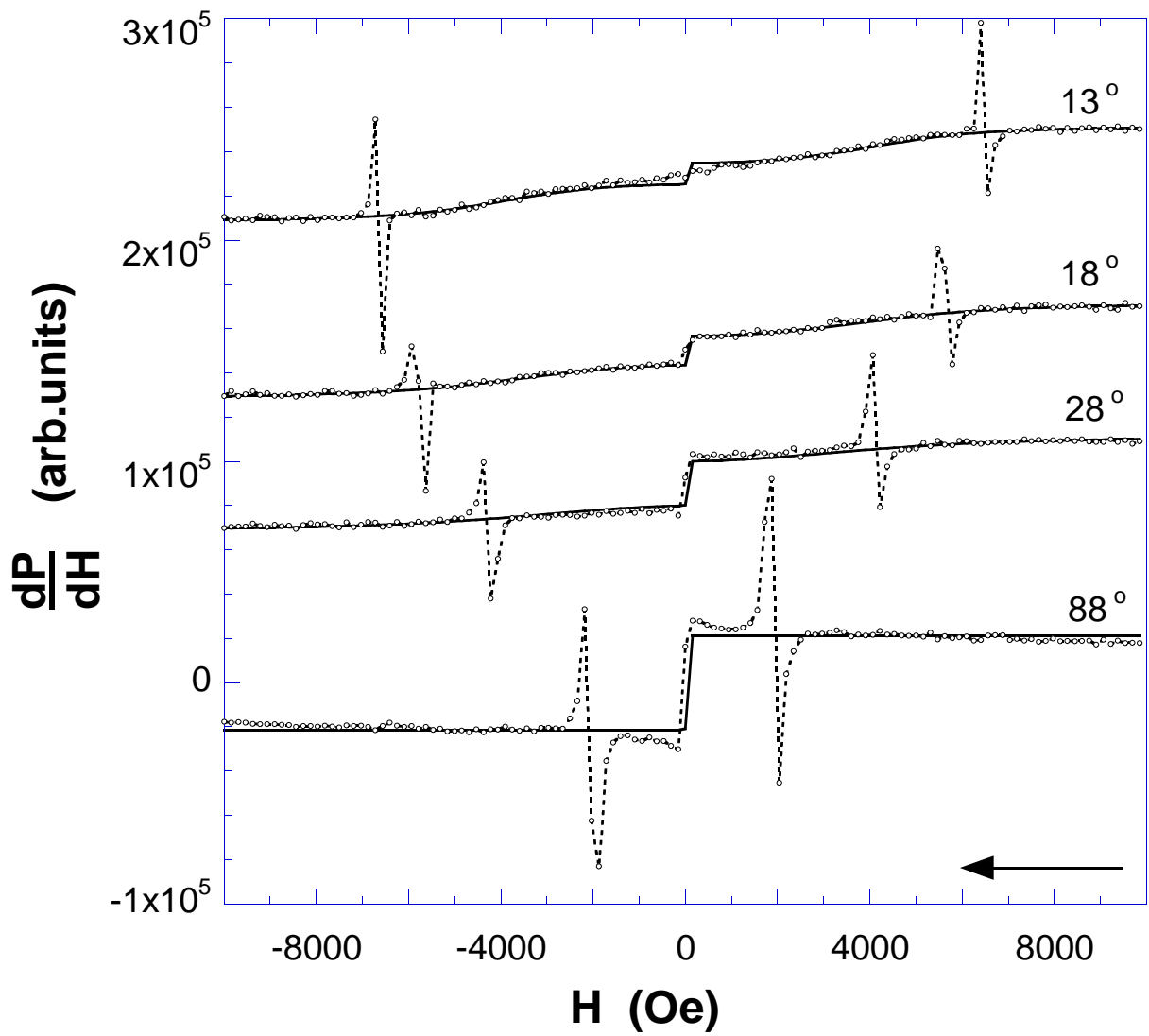
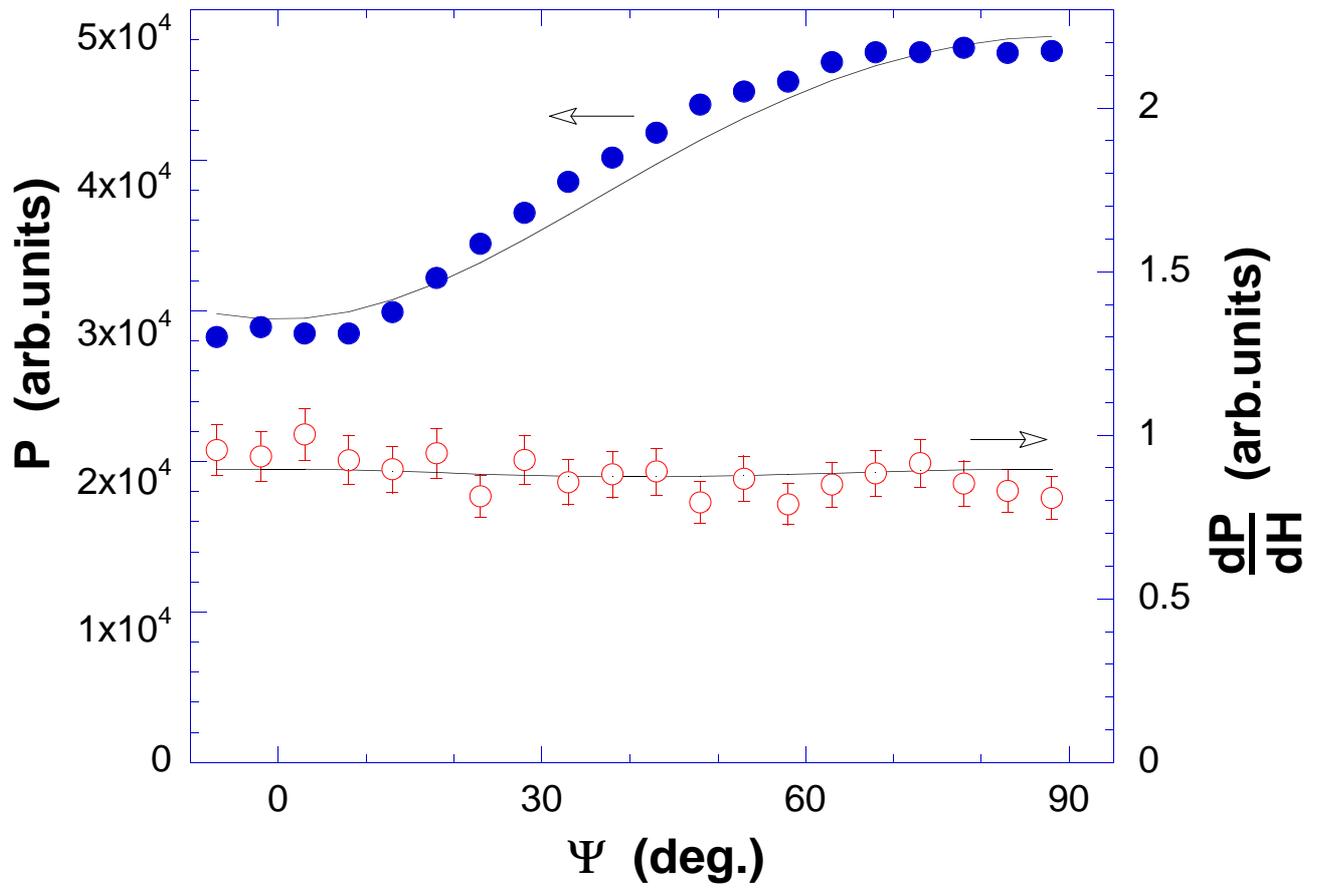
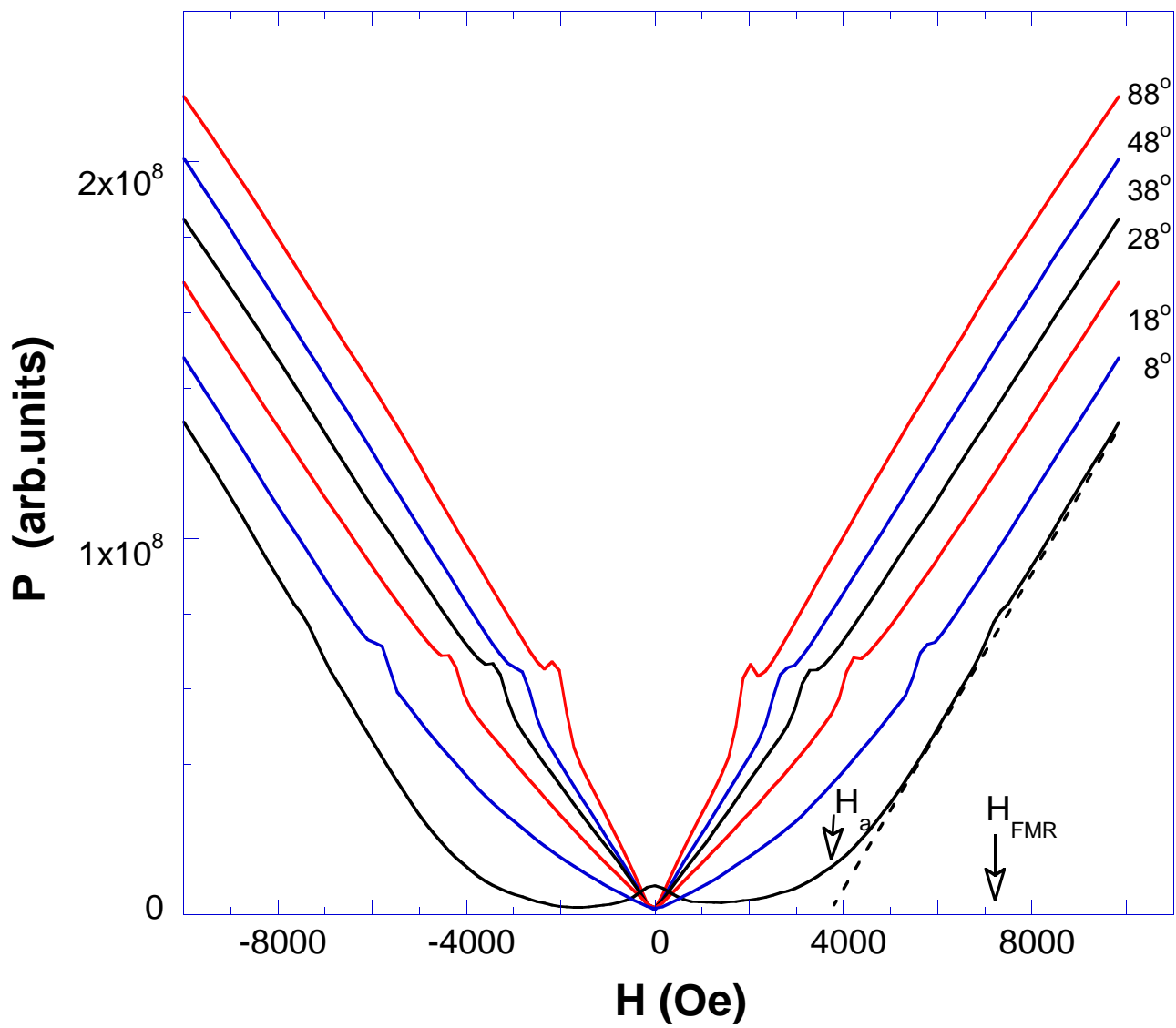
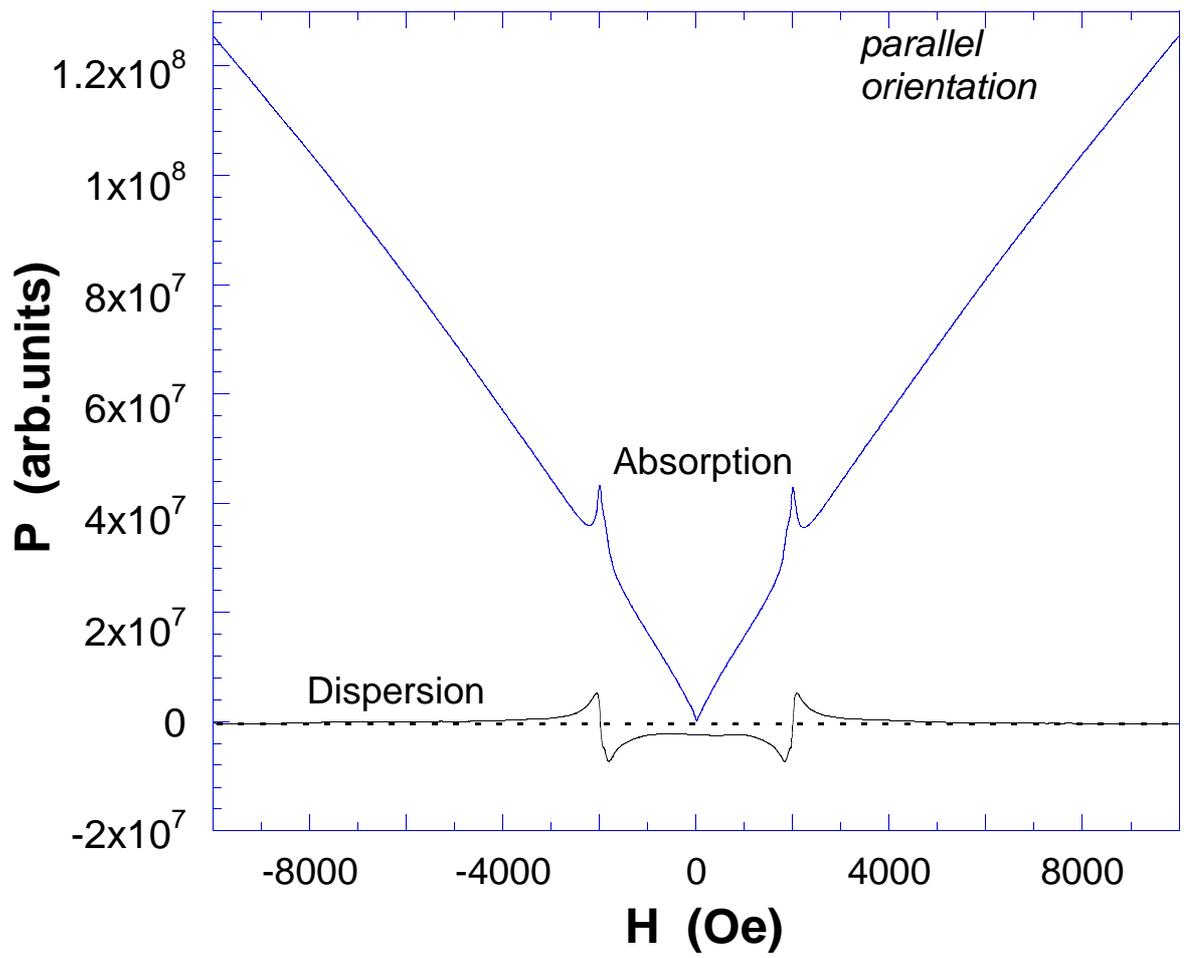
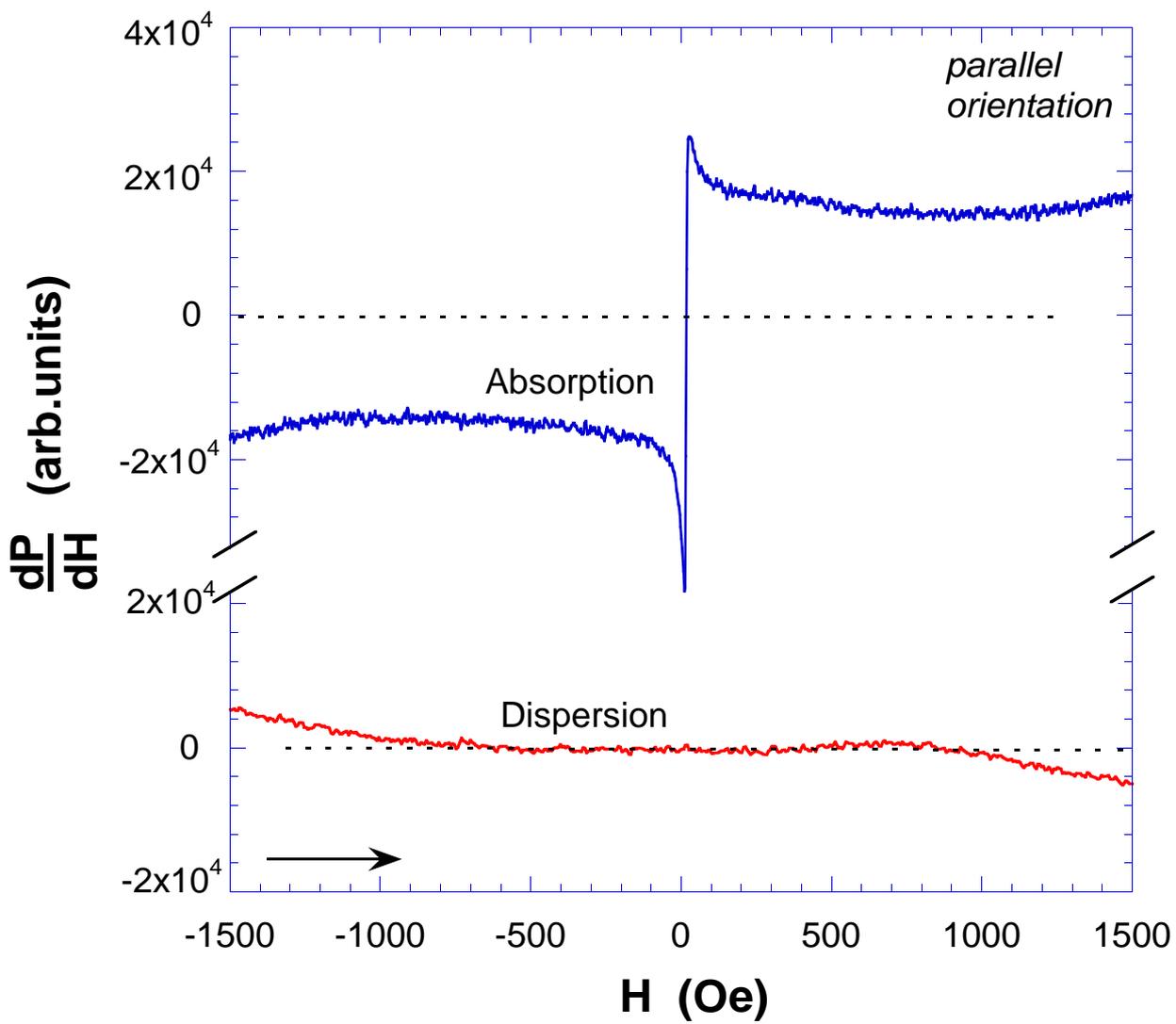
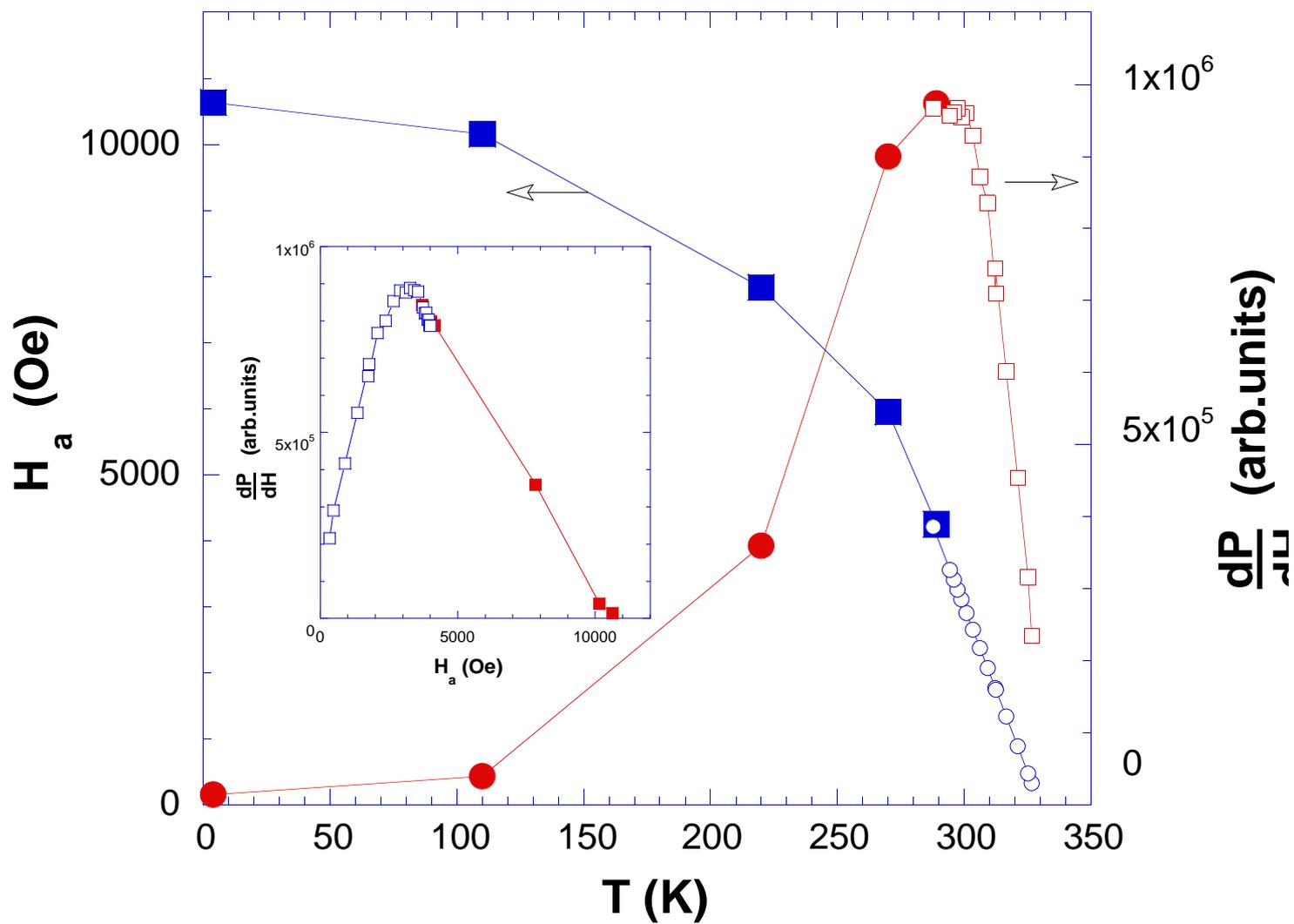
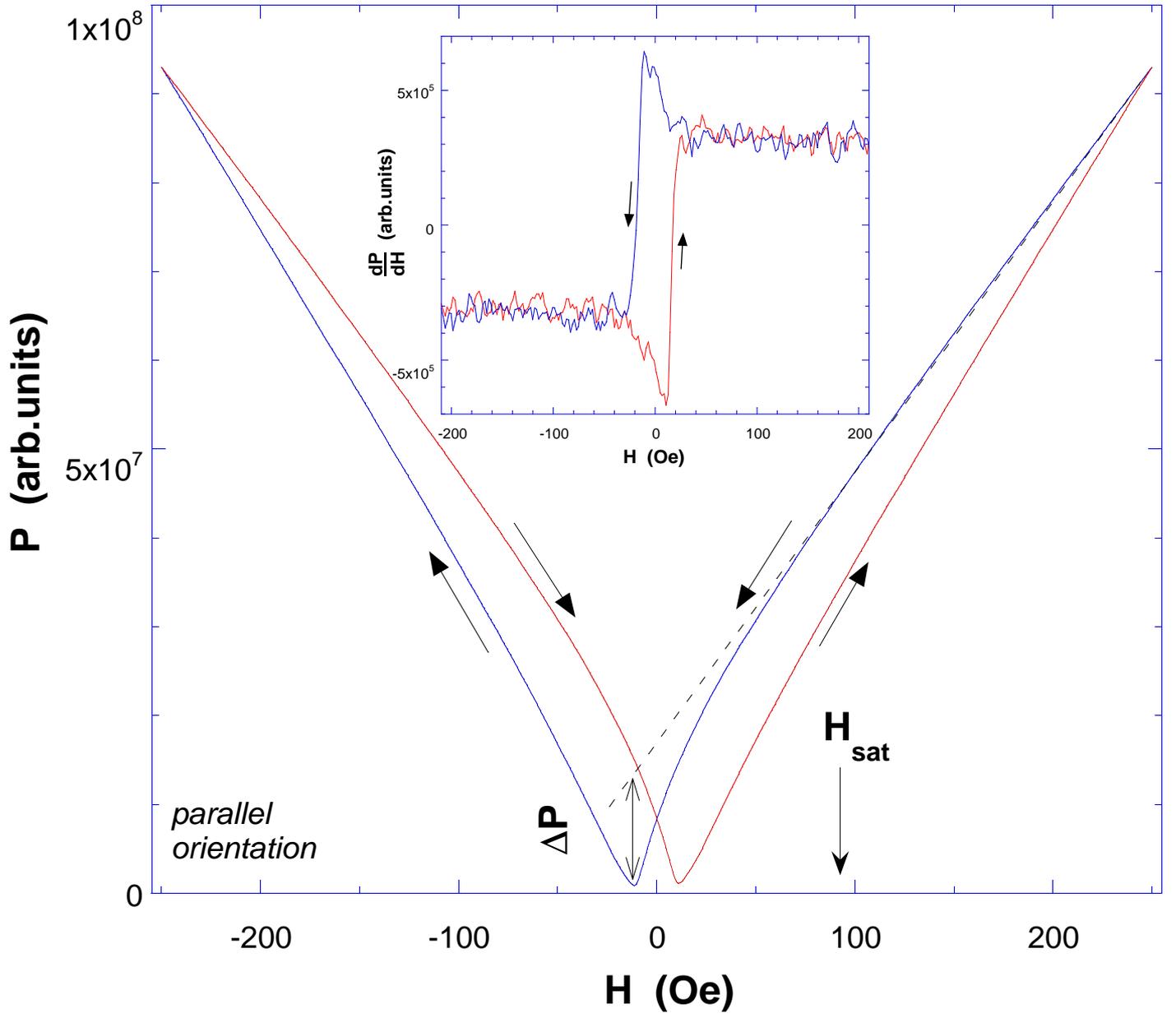
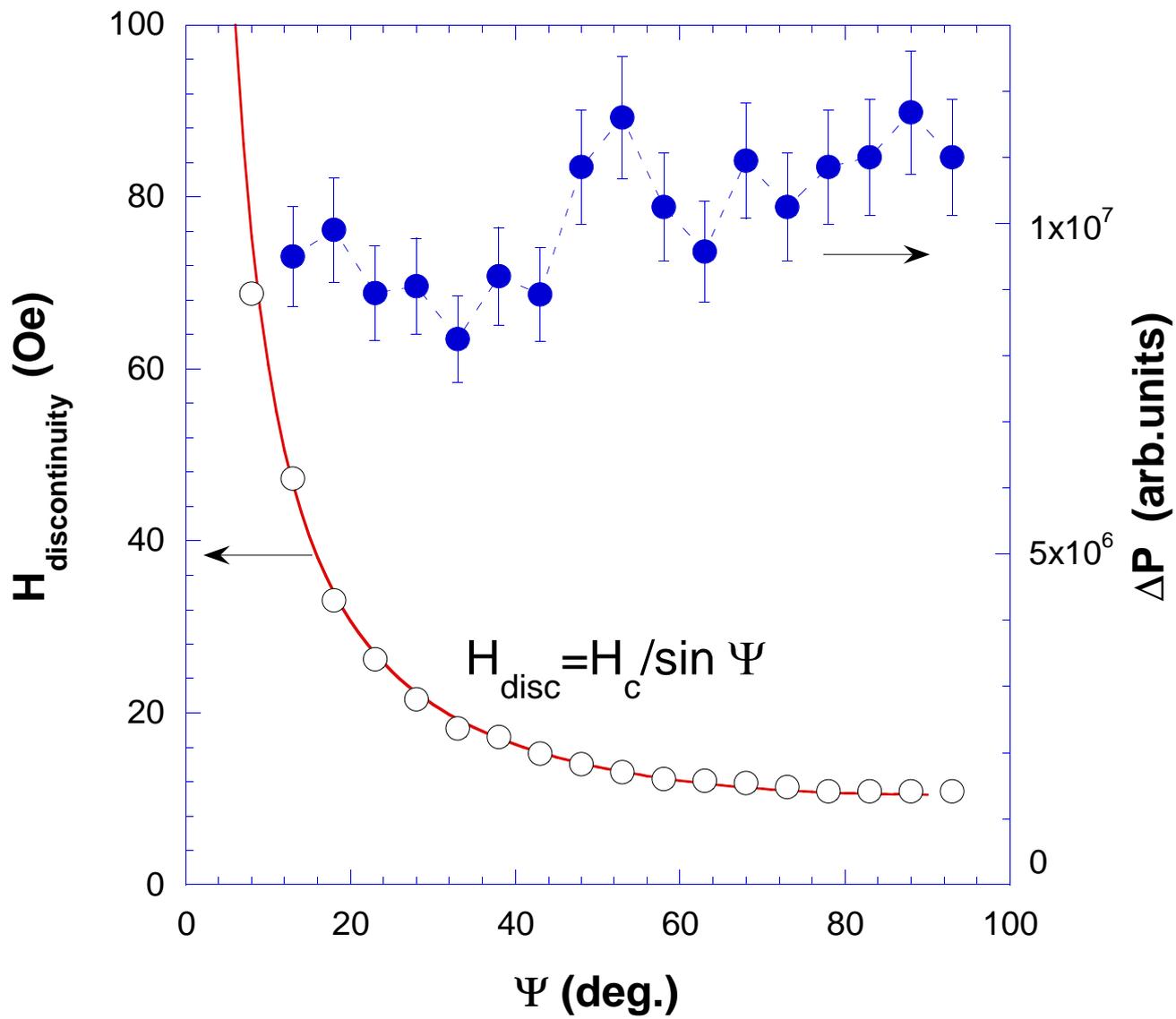
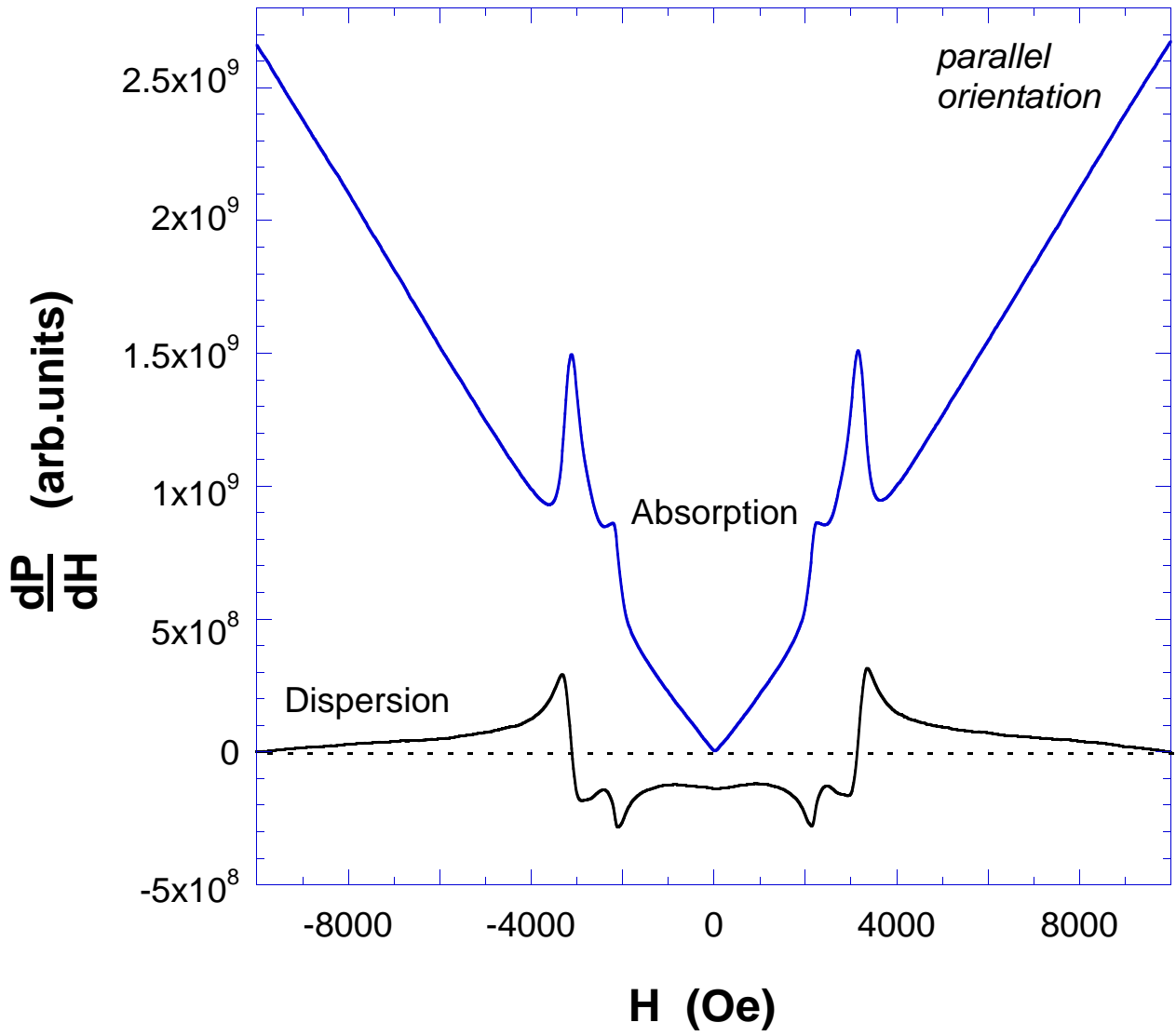
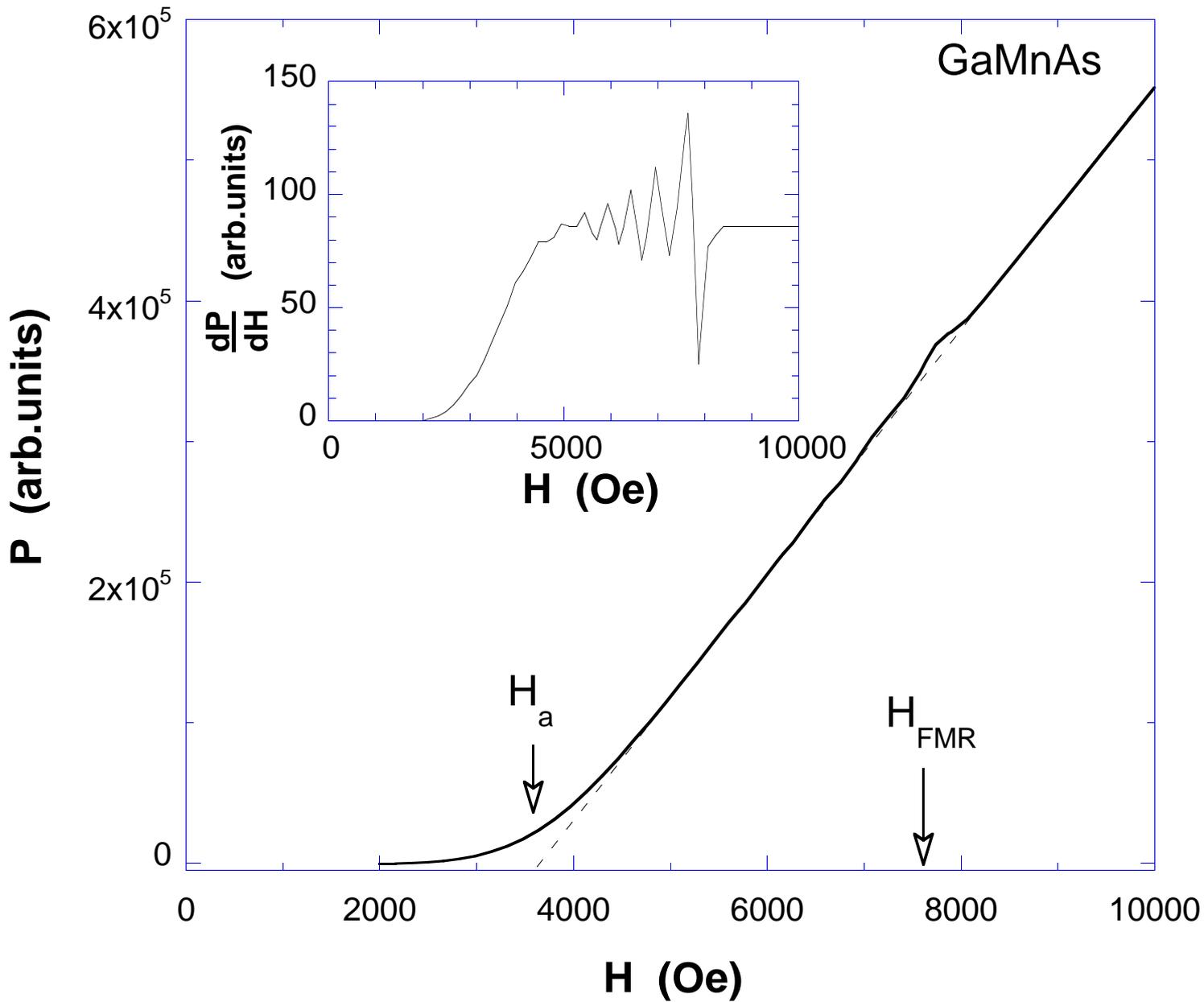
\end{document}